\documentclass[a4paper]{jpconf}
\usepackage{graphicx}
\usepackage[english]{babel}
\usepackage{amsmath,amssymb}

\def\rpv{$R_p \hspace{-1em}/\;\:\hspace{0.2em}$}

\begin{document}

\title{Lepton flavor violation in SUSY left-right symmetric theories}

\author{A.~Vicente}

\address{AHEP Group, Institut de F\'{\i}sica Corpuscular --
  C.S.I.C. \& Universitat de Val{\`e}ncia \\
  Edificio Institutos de Paterna, Apt 22085, E--46071 Valencia, Spain}

\ead{Avelino.Vicente@ific.uv.es}

\begin{abstract}
The seesaw mechanism is the most popular explanation for the smallness of neutrino masses. However, its high scale makes direct tests impossible and only indirect signals at low energies are reachable for collider experiments. One of these indirect links with the high scale is lepton flavor violation (LFV). We discuss LFV decays of sleptons in the context of a SUSY left-right symmetric model that naturally incorporates the seesaw mechanism. This non-minimal embedding of the seesaw leads to observable LFV effects in the right-handed sleptons sector, contrary to minimal models where these are found to be totally negligible. Therefore, LFV observables can be used as a powerful tool to study physics right below the GUT scale.
\end{abstract}

\section{Introduction}

Neutrino physics has been at the forefront of particle phenomenology in the last years. After the discovery of neutrino oscillations, experiments have entered the precision era \cite{Fukuda:1998mi,Collaboration:2007zza,KamLAND2007}, measuring the relevant parameters with increasing accuracy \cite{Schwetz:2008er} and highlighting the need for an explanation of neutrino masses.

The seesaw mechanism is the most popular framework to accommodate neutrino masses \cite{Minkowski:1977sc,seesaw,MohSen,Schechter:1980gr,Cheng:1980qt,Foot:1988aq}. The existence of very heavy fields, right-handed neutrinos in the most common variation, naturally explains the smallness of neutrino masses due to the famous seesaw relation, $m_\nu \sim v_{EW}^2 / M_{heavy}$. However, the structure of the seesaw makes direct tests impossible. The new heavy fields cannot be produced at colliders and therefore at best only indirect tests will be possible.

In addition to the question of neutrino masses, the Standard Model has other theoretical problems that need to be addressed. Several extensions have been built with this purpose, filling the literature with a wide variety of ideas that will be put to experimental test in the coming years. Among the different choices, Supersymmetry (SUSY) is the most popular one because it technically solves the hierarchy problem \cite{Dimopoulos:1981zb} and it has the capability to address many of the other open questions, like the nature of dark matter, gauge coupling unification and radiative symmetry breaking, to mention a few.

One of the key points in supersymmetric model building is R-parity violation/conservation \cite{Fayet:1974pd,Farrar:1978xj}. This discrete symmetry, defined as $R_p = (-1)^{3(B-L)+2s}$ (where $B$ and $L$ stand for baryon and lepton numbers and $s$ for the spin of the particle), is usually imposed to forbid the dangerous baryon and lepton number violating interactions, never seen in nature. If both were simultaneously present, proton decay would be extremely fast, a phenomenological disaster that is prevented by forbidding the R-parity violating (\rpv) couplings. One of the consequences of this new symmetry is the existence of a stable particle, a natural dark matter candidate, and thus its practical importance is beyond any doubt. However, its fundamental origin is totally unknown and in most cases it is introduced by hand in the theory. Compared to the standard model, where baryon and lepton numbers are automatically conserved, this is a step back.

A possible approach to understand the origin of R-parity is to embed the MSSM in an extended model with a gauge group containing a $U(1)_{B-L}$ piece. The original high-energy theory will conserve R-parity due to the gauge symmetry and R-parity will become a remnant subgroup at low energies if the scalar fields responsible for the breaking have even $B-L$ charges.

This picture can be realized in minimal $U(1)_{B-L}$ extensions of the MSSM, see for example \cite{FileviezPerez:2010ek}, or, more ambitiously, in Left-Right (LR) symmetric models \cite{earlyLR}. Apart from the conservation of R-parity this type of models have other motivations. The original motivation was the restoration of parity as an exact symmetry at higher energies \cite{earlyLR}. In addition, it has been shown that they provide technical solutions to the SUSY CP and strong CP problems \cite{Mohapatra:1996vg}, they give an understanding of the $U(1)$ charges and they can be embedded in $SO(10)$ Grand Unified Theories (GUTs).

Here we study some phenomenological aspects of a supersymmetric LR model that leads to R-parity conservation at low energies and incorporates a type-I seesaw mechanism to generate neutrino masses. We will concentrate the discussion on lepton flavor violating signals in slepton decays, such as $\tilde{l}_i \to \tilde{\chi}_1^0 \: l_j$ with $i \neq j$. These signatures have been known to be important in SUSY models for many years \cite{Borzumati:1986qx,Hisano:1995nq,Hisano:1995cp}, and in fact they have been already studied in great detail for minimal seesaw implementations, see for example \cite{Hirsch:2008dy,Hirsch:2008gh,Esteves:2009qr}. This work, however, studies the case of a non-minimal seesaw, what implies new features. In particular, the high-energy restoration of parity enhances the flavor violating effects in the right-handed slepton sector, contrary to the usual expectation. This way, by measuring braching ratios of LFV decays at colliders one can get valuable information on the structure of the high-energy theory.

\section{The model}

\subsection{How to break the LR symmetry}

There are many supersymmetric LR models in the literature. From the pioneering works in the 70s, many extensions and variations have been proposed. In all of them the left-right gauge group $SU(3)_c \times SU(2)_L \times SU(2)_R \times U(1)_{B-L}$ breaks down to the SM gauge group $SU(3)_c \times SU(2)_L \times U(1)_Y$. However, one can choose different representations in the scalar sector, responsible for the breaking of the symmetry, and obtain very different low energy effective theories. In addition, there are several ingredients that cannot be forgotten if one wants to have a consistent framework. Therefore, four requirements will be imposed as guidelines for the choice of the model: (a) Automatic conservation of R-parity, (b) Parity conservation at high energies, (c) Seesaw mechanism, and (d) Cancellation of anomalies.

The first LR models used $SU(2)_R$ doublets to break the gauge symmetry. The non-supersymmetric model proposed in references \cite{earlyLR} introduced two additional scalar doublets $\chi_L$ and $\chi_R$, where $\chi_L \equiv \chi_L(1,2,1,1)$ and $\chi_R \equiv \chi_R(1,1,2,-1)$ under $SU(3)_c \times SU(2)_L \times SU(2)_R \times U(1)_{B-L}$\footnote{Note the duplication in the number of fields. This comes from parity conservation, that implies the same number of $SU(2)_L$ and $SU(2)_R$ charged fields, and anomaly cancellation, that implies that for every group representation with charge $+Q$ under $U(1)_{B-L}$ there must be another one with charge $-Q$. This will be also found in the models discussed below.}. When the neutral component of $\chi_R$ gets a vacuum expectation value (VEV), $\langle \chi_R^0 \rangle \neq 0$, the gauge symmetry is broken down to the SM gauge group and the known low energy phenomenology with broken parity is recovered.

However, such models are not suited for the purpose of this study. The reason comes from the oddness of $\chi_R$ under $U(1)_{B-L}$. In a supersymmetric version of the model, the breaking of the gauge symmetry by $\langle \chi_R^0 \rangle$ also implies the breaking of R-parity. This could be solved by imposing additional discrete symmetries to the model that forbid the dangerous \rpv operators \cite{Malinsky:2005bi}, but this cannot be regarded as automatic R-parity conservation. In addition, there is no seesaw mechanism to generate neutrino masses, and additional superfields would be needed to account for them.

The simplest solution is to break the gauge symmetry by $SU(2)_R$ fields with even charge under $U(1)_{B-L}$. This was in fact proposed in reference \cite{Cvetic:1983su}, where four triplets were added to the MSSM spectrum: $\Delta(1,3,1,2)$, $\Delta^c(1,1,3,-2)$, $\bar{\Delta}(1,3,1,-2)$ and $\bar{\Delta}^c(1,1,3,2)$. In these models, sometimes called MSUSYLR (Minimal Supersymmetric Left-Right), $SU(2)_R \times U(1)_{B-L}$ is broken down to $U(1)_Y$ by the VEVs of the scalar components of $\Delta^c$ and $\bar{\Delta}^c$. When this occurs, a right-handed neutrino mass is generated from the operator $L^c \Delta^c L^c$, leading to a type-I seesaw mechanism. However, the issue of R-parity conservation is not clear. Although one would naively expect that R-parity is automatically conserved due to the even $B-L$ charges of $\Delta^c$ and $\bar{\Delta}^c$, the scalar potential of the theory might not allow to have vanishing sneutrino VEVs \cite{Kuchimanchi:1993jg}, favoring the existence of R-parity breaking minima. For many years, MSUSYLR and its minimal extensions were considered to break R-parity. Nevertheless, the authors of the recent reference \cite{Babu:2008ep} claimed that 1-loop corrections change the picture, allowing for vanishing sneutrino VEVs in the minimum of the potential. Since this is still a controversial issue that relies on not fully calculated loop corrections, we leave this possibility for future studies.

Finally, Aulakh and collaborators \cite{Aulakh:1997ba,Aulakh:1997fq} extended MSUSYLR by the addition of two triplets, $\Omega(1,3,1,0)$ and $\Omega^c(1,1,3,0)$. They showed that the scalar potential offers the new possibility, within large regions of parameter space, of having minima that conserve R-parity while breaking the gauge symmetry in the proper way. Moreover, since the $\Delta$ triplets are part of the spectrum, the seesaw mechanism is present like in MSUSYLR, generating small masses for the light neutrinos. In conclusion, this model fulfills the requirements that we imposed, and thus we will concentrate on it in the following.

\subsection{Model basics}

The matter content of the model is \cite{Aulakh:1997ba,Aulakh:1997fq}

\begin{center}
\begin{tabular}{c c c c c c}
\hline
Superfield & generations & $SU(3)_c$ & $SU(2)_L$ & $SU(2)_R$ & $U(1)_{B-L}$ \\
\hline
$Q$ & 3 & 3 & 2 & 1 & $\frac{1}{3}$ \\
$Q^c$ & 3 & $\bar{3}$ & 1 & 2 & $-\frac{1}{3}$ \\
$L$ & 3 & 1 & 2 & 1 & -1 \\
$L^c$ & 3 & 1 & 1 & 2 & 1 \\
$\Phi$ & 2 & 1 & 2 & 2 & 0 \\
$\Delta$ & 1 & 1 & 3 & 1 & 2 \\
$\bar{\Delta}$ & 1 & 1 & 3 & 1 & -2 \\
$\Delta^c$ & 1 & 1 & 1 & 3 & -2 \\
$\bar{\Delta}^c$ & 1 & 1 & 1 & 3 & 2 \\
$\Omega$ & 1 & 1 & 3 & 1 & 0 \\
$\Omega^c$ & 1 & 1 & 1 & 3 & 0 \\
\hline
\end{tabular}
\end{center}

Here $Q$, $Q^c$, $L$ and $L^c$ contain the quark and lepton superfields of the MSSM with the addition of a right-handed neutrino $\nu^c$. The two $\Phi$ superfields are $SU(2)_L \times SU(2)_R$ bidoublets and contain the usual $H_d$ and $H_u$ MSSM Higgs doublets. Finally, the rest of the superfields are introduced to break the LR symmetry. With these representations, the most general superpotential compatible with the gauge symmetry and parity is

\begin{eqnarray} \label{eq:Wsuppot1}
{\cal W} &=& Y_Q Q \Phi Q^c 
          +  Y_L L \Phi L^c 
          - \frac{\mu}{2} \Phi \Phi
          +  f L \Delta L
          +  f^* L^c \Delta^c L^c \nonumber \\
         &+& a \Delta \Omega \bar{\Delta}
          +  a^* \Delta^c \Omega^c \bar{\Delta}^c
          + \alpha \Omega \Phi \Phi
          +  \alpha^* \Omega^c \Phi \Phi \\
         &+& M_\Delta \Delta \bar{\Delta}
          +  M_\Delta^* \Delta^c \bar{\Delta}^c
          +  M_\Omega \Omega \Omega
          +  M_\Omega^* \Omega^c \Omega^c \nonumber
\end{eqnarray}

Family and gauge indices have been omitted in equation \eqref{eq:Wsuppot1}. $Y_Q$ and $Y_L$ are the usual quark and lepton Yukawa couplings. However, note that $Y_Q Q \Phi Q^c \equiv Y_Q^\alpha Q \Phi_\alpha Q^c$ and $Y_L L \Phi L^c \equiv Y_L^\alpha L \Phi_\alpha L^c$, with $\alpha = 1,2$, and thus there are four $3 \times 3$ Yukawa matrices. Conservation of parity implies that they must be symmetric. $f$ is a $3 \times 3$ complex symmetric matrix, whereas $\alpha$ is a $2 \times 2$ antisymmetric matrix, and thus it only contains one complex parameter, $\alpha_{12}$.

The breaking of the LR gauge group to the MSSM gauge group happens in two steps.

\begin{displaymath}
SU(2)_R \times U(1)_{B-L} \quad \longrightarrow \quad U(1)_R \times U(1)_{B-L} \quad \longrightarrow \quad U(1)_Y
\end{displaymath}

The first step is due to the VEV of the $\Omega^{c \: 0}$ field $\langle \Omega^{c \: 0} \rangle = \frac{v_R}{\sqrt{2}}$, which breaks $SU(2)_R$. However, note that, since $T_{3R} (\Omega^{c \: 0}) = 0$ there is a $U(1)_R$ symmetry left over. Next, the group $U(1)_R \times U(1)_{B-L}$ is broken by $\langle \Delta^{c \: 0} \rangle = \frac{v_{BL}}{\sqrt{2}}$ and $\langle \bar{\Delta}^{c \: 0} \rangle = \frac{\bar{v}_{BL}}{\sqrt{2}}$. The remaining symmetry is $U(1)_Y$, with hypercharge defined as $Y = I_{3R} + \frac{B-L}{2}$.

Note that, since the tadpole equations do not link $\Omega^c$, $\Delta^c$ and $\bar{\Delta}^c$ with their left-handed counterparts, the left-handed triplets can be taken to have vanishing vevs \cite{Aulakh:1997ba}.

Although a hierarchy between the two breaking scales may exist, $v_{BL} \ll v_R$, one cannot neglect the effects of the second breaking stage on the first one. The tadpole equations mix them, and only through the contribution of the $\Delta^c-\bar{\Delta}^c$ fields one can understand a non-vanishing $v_R$ VEV. In fact, there is an inverse hierarchy between the VEVs and the superpotential masses $M_\Delta$, $M_\Omega$, given by

\begin{equation} \label{tadpolesol}
v_R = \frac{2 M_\Delta}{a} \qquad v_{BL} = \frac{2}{a} (2 M_\Delta M_\Omega)^{1/2}
\end{equation}

And so, for $v_{BL} \ll v_R$ one needs $M_\Delta \gg M_\Omega$ \cite{Aulakh:1997ba}.

\section{Slepton decays and LFV}

Lepton flavor violation is a well known indirect test of the seesaw mechanism \cite{Borzumati:1986qx,Hisano:1995nq,Hisano:1995cp}. Assuming flavor-blind soft SUSY breaking terms at some high-energy scale, the RGE running down to the SUSY scale generates non-zero off-diagonal entries in the slepton soft squared masses. These flavor violating entries are connected to the effective neutrino mass matrix and thus, by making some assumptions, one can find testable relations. Moreover, they induce LFV decays, such as $l_i \to l_j \gamma$ and $\tilde{l}_i \to \tilde{\chi}_1^0 \: l_j$ with $i \neq j$. By studying these decays one can set important constraints and get valuable information on the underlying theory.

In all cases studied in the literature, based on minimal seesaw models, the off-diagonal entries of the soft masses of right-handed sleptons get negligible contributions from the RGE running and thus one expects no visible signal of LFV from the right-handed sector at the LHC. However, in LR models the gauge symmetry makes left- and right-handed sectors behave the same, and then the LFV violating entries of the soft squared masses in the right-handed sector must contain non-negligible contributions. This novel LHC signal would point to a high-energy LR symmetry in a very clean way.

That is the main result of this work. As we will show below, LFV in the right-handed lepton/slepton sector can be observable in this model. In order to prove it we perform a numerical calculation using the code SPheno \cite{Porod:2003um}, including 2-loop RGEs and the corresponding 1-loop threshold corrections at the intermediate scales. The Yukawa parameters $Y_L$ are fixed in order to correctly reproduce neutrino oscillation data. Finally, all analytical computations have been done with the help of the Mathematica package Sarah \cite{sarah}.

\begin{figure}
\begin{center}
\vspace{5mm}
\includegraphics[width=0.49\textwidth]{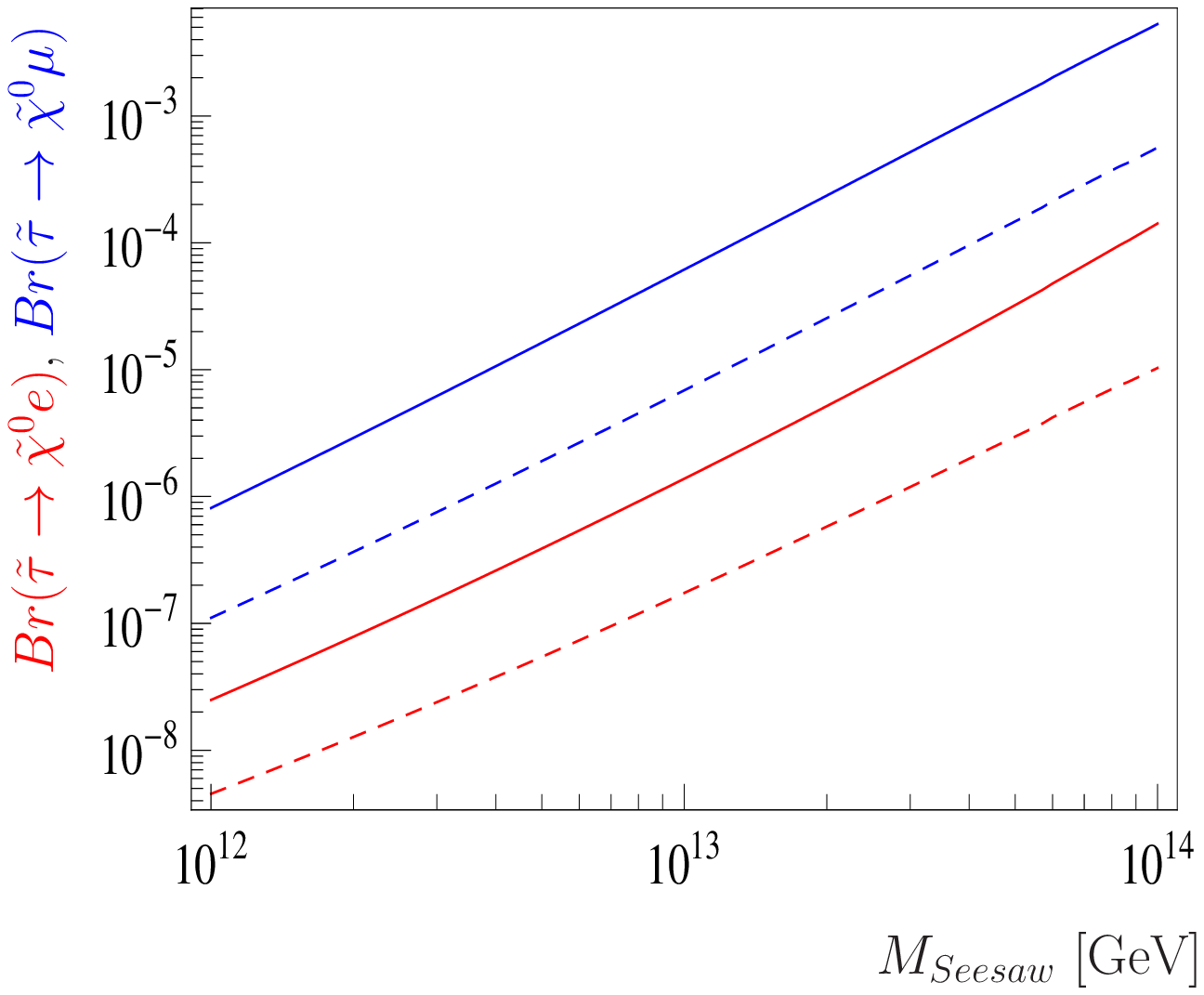}
\end{center}
\vspace{-5mm}
\caption{$Br(\tilde{\tau}_i \to \tilde{\chi}_1^0 \: e)$ and $Br(\tilde{\tau}_i \to \tilde{\chi}_1^0 \: \mu)$ as a function of the seesaw scale, defined as $M_{Seesaw} \equiv f v_{BL}$, for the parameter choice $v_{BL} = 10^{15}$ GeV and $v_R = 5 \cdot 10^{15}$ GeV. The dashed lines correspond to $\tau_1 \simeq \tau_R$, whereas the solid ones correspond to $\tau_2 \simeq \tau_L$. The mSUGRA parameters have been taken as in the SPS3 benchmark point \cite{Allanach:2002nj} and the right-handed neutrino spectrum has been assumed to be degenerate, $M_{R i} = M_{Seesaw}$.}
\label{fig:FVvsSeesaw}
\end{figure}

Figure \ref{fig:FVvsSeesaw} shows $Br(\tilde{\tau}_i \to \tilde{\chi}_1^0 \: e)$ and $Br(\tilde{\tau}_i \to \tilde{\chi}_1^0 \: \mu)$ as a function of the seesaw scale, defined as $M_{Seesaw} \equiv f v_{BL}$. The dependence on the seesaw scale is clearly understood from the seesaw formula. This implies that larger $M_{Seesaw}$ requires larger Yukawa parameters in order to fit neutrino masses which, in turn, leads to larger flavor violating terms due to RGE running. 

Furthermore, figure \ref{fig:FVvsSeesaw} also shows that right-handed staus can also have LFV decays with observable rates.  This is the main novelty in this model. One can see that for large $M_{Seesaw}$ values, around $10^{13} - 10^{14}$ GeV, the rates for LFV are measurable for both left- and right-handed staus. See references \cite{Andreev:2006sd,delAguila:2008iz} for the LHC discovery potential in the search for LFV.

The previous result can be understood by using analytical approximations for the slepton soft squared masses. The running from the GUT scale to the SUSY scale generates off-diagonal entries $\Delta m^2$ in both left- and right-handed slepton soft masses. In the first step, from the GUT scale to the $v_R$ scale they can be written in leading-log approximation as \cite{Chao:2007ye}

\begin{eqnarray}
\Delta m_L^2 &=& - \frac{1}{4 \pi^2} \left( 3 f f^\dagger + Y_L^{(k)} Y_L^{(k) \: \dagger} \right) (3 m_0^2 + A_0^2) \ln \left( \frac{m_{GUT}}{v_R} \right) \label{rge1}\\
\Delta m_{L^c}^2 &=& - \frac{1}{4 \pi^2} \left( 3 f^\dagger f + Y_L^{(k) \: \dagger} Y_L^{(k)} \right) (3 m_0^2 + A_0^2) \ln \left( \frac{m_{GUT}}{v_R} \right) \label{rge2}
\end{eqnarray}

After parity breaking at $v_R$ the Yukawa coupling $Y_L$ splits into $Y_e$, the charged lepton Yukawa, and $Y_\nu$, the neutrino Yukawa. The later contributes to LFV entries in the running down to the $v_{BL}$ scale

\begin{eqnarray}
\Delta m_L^2 &=& - \frac{1}{8 \pi^2} Y_\nu Y_\nu^\dagger (3 m_0^2 + A_0^2) \ln \left( \frac{v_R}{v_{BL}} \right) \\
\Delta m_{\tilde{e}^c}^2 &=& 0
\end{eqnarray}

Finally, from $v_{BL}$ to the SUSY scale one recovers the MSSM RGEs, which do not add any flavor violating effect.

This short discussion shows an important consequence of the symmetry breaking pattern. From the GUT scale to the $v_R$ scale parity is conserved and the magnitude of the LFV entries in the left- and right-handed sectors is the same, see eqs. \eqref{rge1} and \eqref{rge2}. However, from $v_R$ to $v_{BL}$ only the left-handed ones keep running, and thus one expects larger flavor violation in this sector. Moreover, if the difference between $v_R$ and $v_{BL}$ is increased, the difference between the LFV entries in the L and R sectors gets increased as well.

\begin{figure}
\begin{center}
\vspace{5mm}
\includegraphics[width=0.49\textwidth]{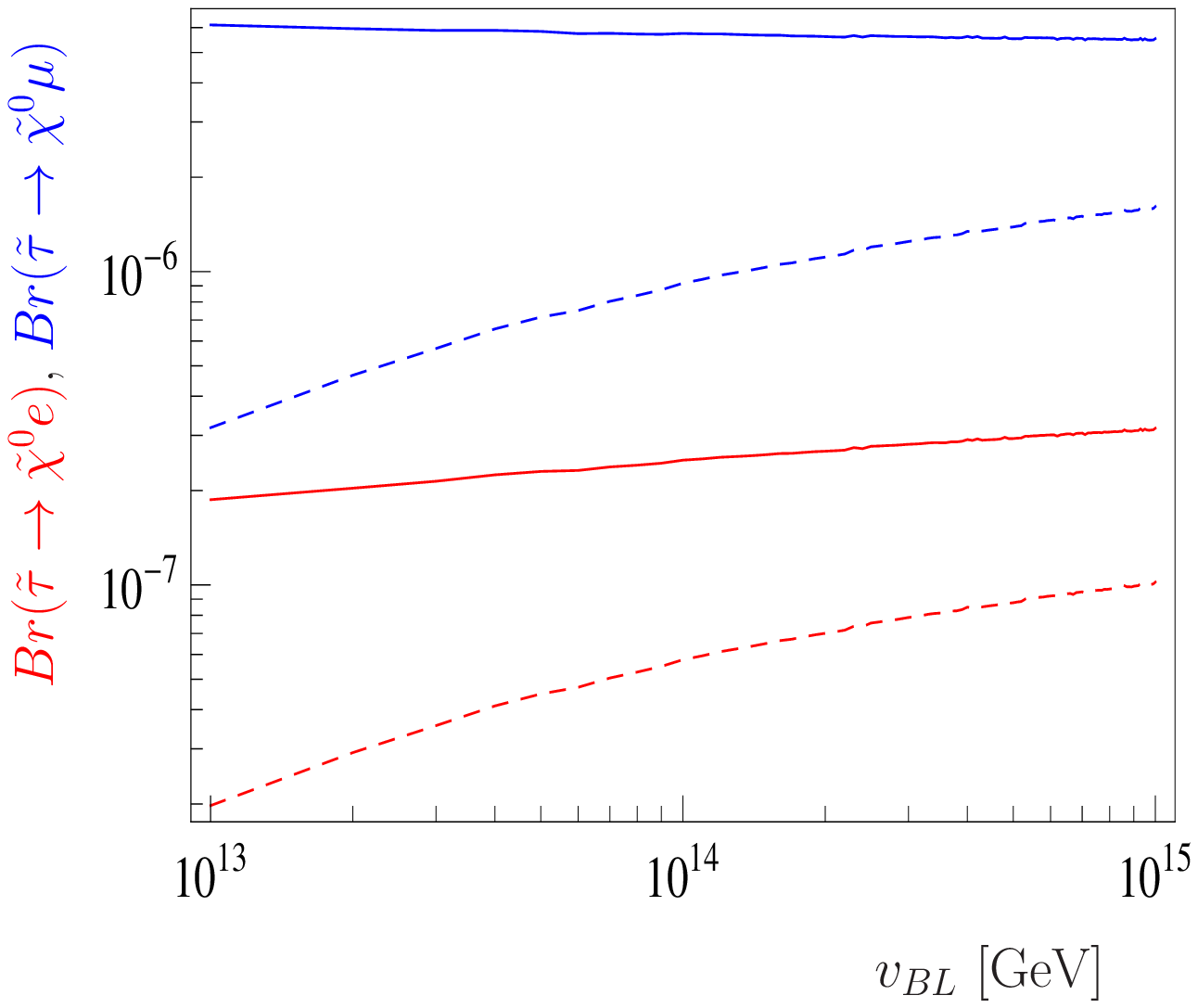}
\includegraphics[width=0.49\textwidth]{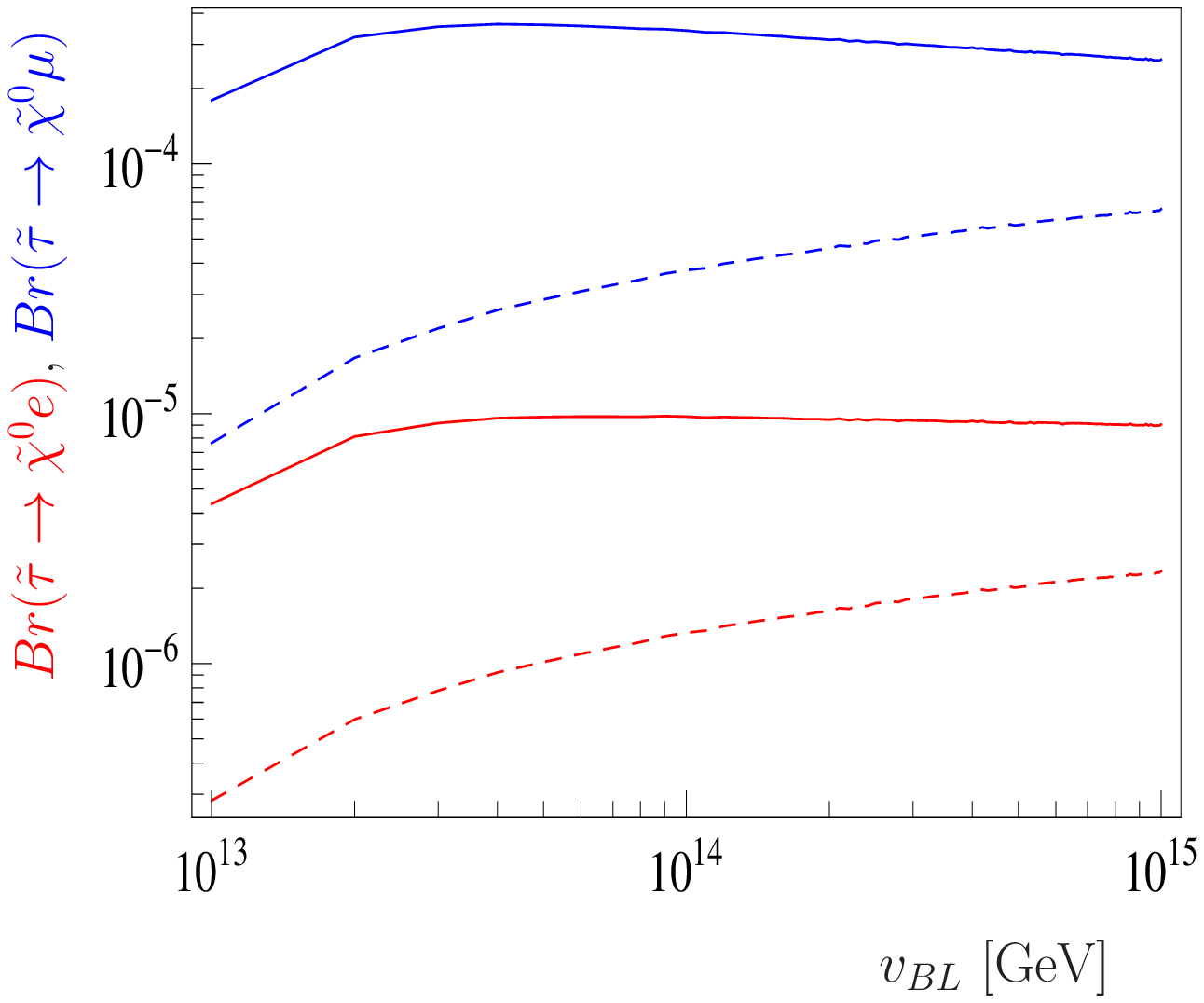}
\end{center}
\vspace{-5mm}
\caption{$Br(\tilde{\tau}_i \to \tilde{\chi}_1^0 \: e)$ and $Br(\tilde{\tau}_i \to \tilde{\chi}_1^0 \: \mu)$ as a function of $v_{BL}$, for a fixed $v_R = 3 \cdot 10^{15}$ GeV. To the left, $M_{Seesaw} = 10^{12}$ GeV, whereas to the right $M_{Seesaw} = 10^{13}$ GeV. The dashed lines correspond to $\tau_1 \simeq \tau_R$, whereas the solid ones correspond to $\tau_2 \simeq \tau_L$. The mSUGRA parameters have been taken as in the SPS3 benchmark point \cite{Allanach:2002nj} and the right-handed neutrino spectrum has been assumed to be degenerate, $M_{R i} = M_{Seesaw}$.}
\label{fig:difLR}
\end{figure}

This is shown in figure \ref{fig:difLR}, which shows branching ratios for the LFV decays of the staus as a function of $v_{BL}$ for a fixed value of $v_R = 3 \cdot 10^{15}$ GeV. The theoretical expectation is obtained: the difference between $Br(\tilde{\tau}_L)$ and $Br(\tilde{\tau}_R)$ strongly depends on the difference between $v_R$ and $v_{BL}$.

\begin{figure}
\begin{center}
\vspace{5mm}
\includegraphics[width=0.49\textwidth]{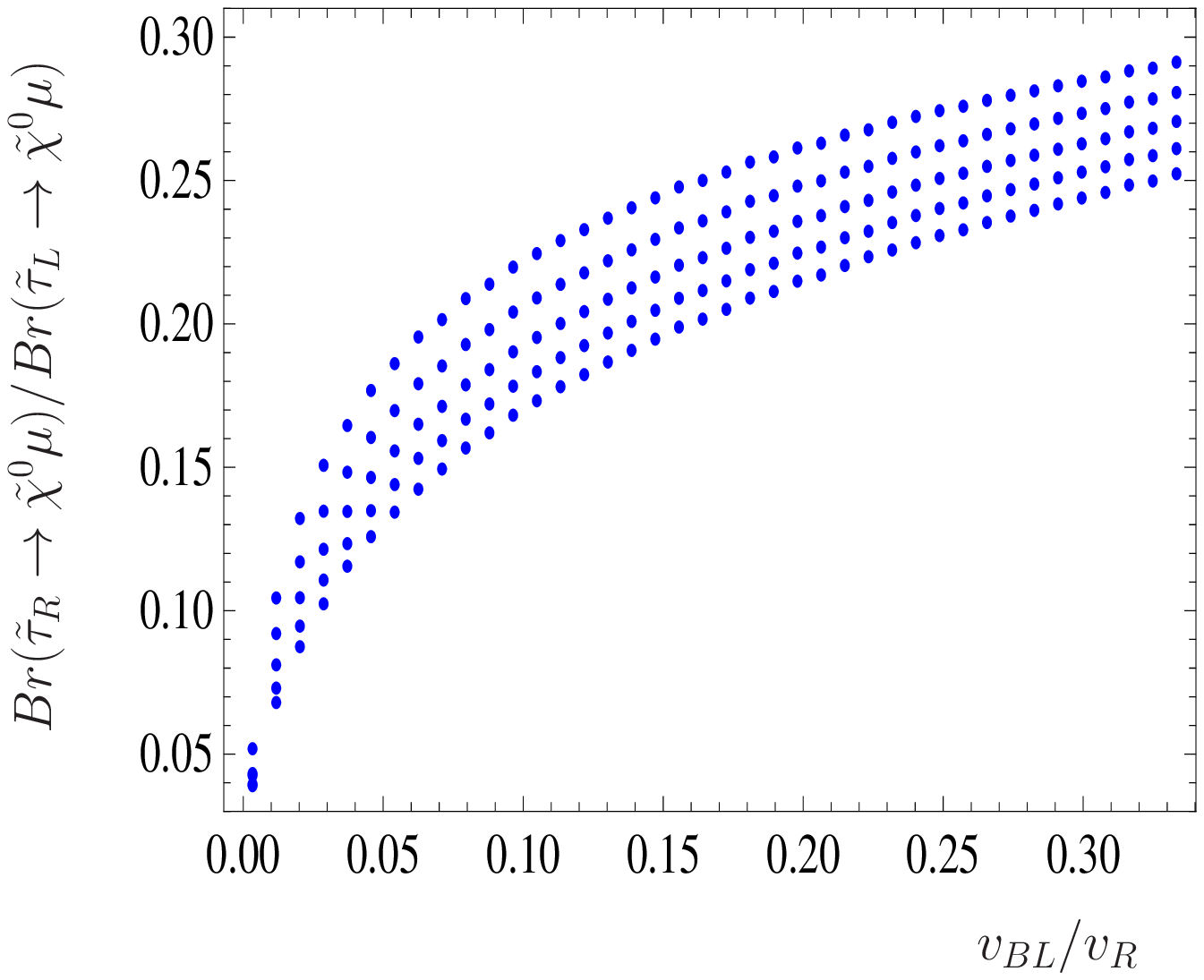}
\end{center}
\vspace{-5mm}
\caption{$Br(\tilde{\tau}_R \to \tilde{\chi}_1^0 \: \mu) / Br(\tilde{\tau}_L \to \tilde{\chi}_1^0 \: \mu)$ as a function of $v_{BL} / v_R$. The seesaw scale $M_{Seesaw}$ takes values in the range $[10^{12},10^{13}]$ GeV. The rest of the parameters have been chosen as in figure \ref{fig:difLR}.}
\label{fig:compLR}
\end{figure}

The question arises whether one can determine the ratio $v_{BL}/v_R$ by measuring both $Br(\tilde{\tau}_L)$ and $Br(\tilde{\tau}_R)$ at the LHC. This is answered in figure \ref{fig:compLR} where the ratio $Br(\tilde{\tau}_R \to \tilde{\chi}_1^0 \: \mu) / Br(\tilde{\tau}_L \to \tilde{\chi}_1^0 \: \mu)$ is plotted as a function of $v_{BL} / v_R$. A measurement of both branching ratios would allow to constrain the ratio $v_{BL} / v_R$. However, there is a slight dependence on other important quantities, such as $m_{GUT}$ and $M_{Seesaw}$. This implies that more experimental information will be needed in order to set reliable constraints.

\section{Summary and conclusions}

Neutrino masses and R-parity conservation are two issues not addressed in the MSSM. On the one hand, the MSSM does not provide an explanation for the observation of neutrino oscillations and the subsequent non-zero neutrino masses. These experimental results require the introduction of a mechanism that can explain the smallness of the neutrino masses, being the seesaw mechanism the most popular choice. On the other hand, R-parity is introduced in the MSSM as an ad-hoc symmetry, without any theoretical motivation. It is therefore interesting to study extended symmetry groups that can lead to R-parity conservation at low energies.

In this work we have studied some phenomenological aspects of a supersymmetric Left-Right model which automatically conserves R-parity and contains the seesaw mechanism to generate neutrino masses. We have found that, contrary to minimal realizations of the seesaw, large lepton flavor violating effects are obtained both in the left- and right-handed slepton sectors. This is a useful signature that allows us to get additional information on the high energy regime and clearly points to an underlying left-right symmetry.

In particular, we have shown that observables like $Br(\tilde{\tau}_R \to \tilde{\chi}_1^0 \: l)$ can get strong deviations from the standard seesaw picture, allowing us to constrain the parameters of the high energy theory and get a hint on its structure. Furthermore, there are other observables which are also very sensitive to new right-handed flavor violation. Examples are slepton mass splittings and the polarization of the outgoing electrons in $\mu \to e \gamma$. We plan to address these issues in a future publication \cite{future}.

\ack

This talk was based on work in collaboration with J. Esteves, M. Hirsch, W. Porod, J. C. Romao and F. Staub, and is supported by the Spanish MICINN under grants FPA2008-00319/FPA, FPA2008-04002-E and MULTIDARK Consolider CAD2009-00064, by Prometeo/2009/091 and by the EU grant UNILHC PITN-GA-2009-237920. A.V. thanks the Generalitat Valenciana for financial support.

\end{document}